\documentclass[journal, final, twocolumn,10pt]{IEEEtran}  

\usepackage{cite}
\usepackage{graphicx}
\usepackage{subfigure}
\usepackage{mathrsfs}
\usepackage{amsmath}
\usepackage{amsfonts}
\usepackage{amssymb}
\usepackage{algorithm2e}
\usepackage{paralist}
\usepackage{color}
\usepackage{epsfig}
\usepackage{pst-all}
\usepackage{pstricks-add}

\newtheorem{theorem}{{Theorem}}
\newtheorem{lemma}[theorem]{{Lemma}}

\newtheorem{definition}[theorem]{{Definition}}

\newcommand{\qed}{\hspace*{\fill} $\Box$ \\}

\begin{document}

\title{Decentralized Coding Algorithms for Distributed Storage in Wireless
 Sensor Networks}

\author{Zhenning~Kong,~\IEEEmembership{Student Member,~IEEE,}
        Salah~A.~Aly,~\IEEEmembership{Member,~IEEE,}
       Emina~Soljanin,~\IEEEmembership{Senior Member,~IEEE,}
\thanks{Manuscript received January 29, 2009; revised August 7, 2009.}
\thanks{Parts of this work were presented in IPSN'08 and ISIT'08 conferences.}
\thanks{Z.~Kong is with the Department of Electrical Engineering, Yale University, New Haven, CT 06520, USA, (email: zhenning.kong@yale.edu).}
\thanks{S.~A.~Aly is with the Department of Computer Science, Texas A\&M University, College Station, TX 77843, USA, (email: salah@cs.tamu.ed).}
\thanks{E.~Soljanin is with Bell Laboratories, Alcatel-Lucent,
Murray Hill, NJ 07974, USA, (email: emina@lucent.com).}}

\markboth{\emph{IEEE Journal on Selected Areas in Communications,} to appear in 2010}
{\emph{IEEE Journal on Selected Areas in Communications,} to appear in 2009}

\maketitle

\begin{abstract}

We consider large-scale wireless sensor networks with $n$ nodes, out of which $k$
are in possession, ({\it e.g.,} have sensed or collected in some
other way) $k$ information packets. In the scenarios in which
network nodes are vulnerable because of, for example, limited energy
or a hostile environment, it is desirable to disseminate the
acquired information throughout the network so that each of the $n$
nodes stores one (possibly coded) packet so that the original $k$
source packets can be recovered, locally and in a computationally simple
way from any $k(1 + \epsilon)$ nodes for some small $\epsilon > 0$.
We develop decentralized Fountain codes based algorithms to solve this
problem. Unlike all previously developed schemes, our algorithms are
truly distributed, that is, nodes do not know $n$, $k$ or
connectivity in the network, except in their own neighborhoods, and
they do not maintain any routing tables.
\end{abstract}

\section{Introduction}

Wireless sensor networks consist of small devices (sensors) with limited resources (e.g.,
low CPU power, small bandwidth, limited battery and memory). They are
mainly used to monitor and detect objects, fires, temperatures, floods, and other phenomena~\cite{RaSiZn04},
often in challenging environments where human involvement is limited. Consequently,
data acquired by sensors may have short lifetime, and any processing of such data within
the network should have low complexity and power consumption~\cite{RaSiZn04}.

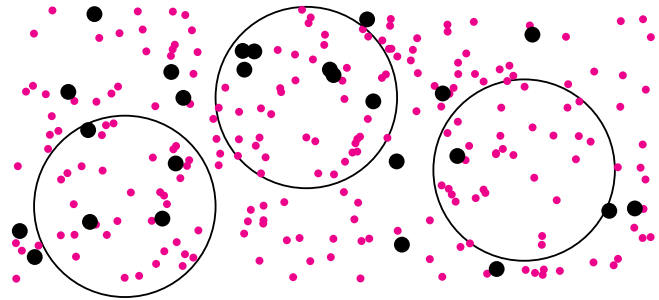
\begin{figure}[hbt]
\psset{unit=1.9in} \psset{linewidth=0.7pt}
\begin{pspicture}(0,0)(1.8,.75)
\pscircle(.3,.2){.25}\pscircle(.8,.5){.25}\pscircle(1.4,.3){.25}
\psRandom[dotsize=3pt,randomPoints=225,linecolor=magenta](1.75,.75){}
\psRandom[dotsize=6pt,randomPoints=25](1.75,.75){}
\end{pspicture}
\caption{A sensor network has 25 sensors (big dots) monitoring an area and 225 storage nodes
(small dots). A good distributed storage algorithm should enable us to recover the original 25
source packets from any 25+ nodes (e.g., the set of nodes within any one of the three
illustrated circular regions).} \label{fig:SensorNetworks}
\end{figure}

Consider a wireless sensor network with $n$ sensors, where $k$ sensors collect(sense) independent information.
Because of the network vulnerability and/or inaccessibility, it is desirable to
disseminate the acquired information throughout the network so that each of the $n$ nodes stores
one (possibly coded) packet and the original $k$ source packets can be recovered in a
computationally simple way from any $k(1+\epsilon)$ of nodes for some small $\epsilon>0$. Two such
scenarios are of particular practical interest: to have the information acquired by the $k$
sensors recoverable (1) locally from any neighborhood containing $k(1+\epsilon)$ nodes or (2) from the
last $k(1+\epsilon)$ surviving nodes. Fig.~\ref{fig:SensorNetworks} illustrates
such an example.

Many algorithms have been proposed to solve related distributed storage
problems using coding with either centralized or mostly decentralized control.
Reed-Solomon based schemes have
been proposed in \cite{weatherspoon02,DiPrRa05,DiPrRa06-1,pitkanen06} and Low-Density Parity Check codes based schemes in \cite{huang05,plank05,plank07}, and references therein.

Fountain codes have also been considered because they are rateless and because
of their coding efficiency and low complexity.
In~\cite{DiPrRa06-2} Dimakis~\emph{el al.} proposed a decentralized implementation of
Fountain codes using fast random walks to disseminate source data to the storage nodes
and geographic routing over a grid, which requires every node to know its location.
In~\cite{LiLiLi07-1}, Lin~\emph{et al.} proposed a solution employing random walks with
stops, and used the Metropolis algorithm to specify transition probabilities of the random walks.

In another line of work, Kamra \emph{et al.} in~\cite{KaMiFeRu06} proposed a novel technique
called growth coding to increase data persistence in wireless sensor networks, that is,
the amount of information that can be recovered at any storage node at any time period
whenever there is a failure in some other nodes. In~\cite{LiLiLi07-2}, Lin~\emph{et al.} described
how to differentiate data persistence using random linear codes.
{Network coding has also been considered for distributed storage in various networks
scenarios~\cite{jiang06,wang06,AcDeMeKo05,DiGoWaRa07,MuWiRoZo07}.}

All previous work  assumes some access to global
information, for example, the total numbers of nodes and sources, which, for
large-scale wireless sensor networks, may not be easily obtained
or updated by each individual sensor. By contrast, the algorithms
proposed in this paper require no global information. For example,
in~\cite{LiLiLi07-1}, the knowledge of the total
number of sensors $n$ and the number of sources $k$ is required to calculate the number
of random walks that each source has to initiate, and the probability of trapping data at each
sensor. The knowledge of the maximum node degree (i.e., the maximum number of node neighbors) of the
graph is also required to perform the Metropolis algorithm.
Furthermore, the algorithms proposed in~\cite{LiLiLi07-1} request each
sensor to perform encoding only after receiving enough source packets. This demands each
sensor to maintain a large temporary memory buffer, which may not be practical in
real sensor networks.

In this paper, we propose two new algorithms to solve the distributed storage problem for
large-scale wireless sensor networks: LT-Codes based distributed
storage (LTCDS) algorithm and Raptor Codes based distributed storage (RCDS) algorithm.
Both algorithms employ simple random walks. Unlike all previously developed schemes, both LTCDS
and RCDS algorithms are truly distributed. That is, except for their own neighborhoods, sensors do
not need to know any global information, e.g., the total number of sensors $n$, the number of
sources $k$, or routing tables. Moreover, in both algorithms, instead of waiting until all the
necessary source packets have been collected to perform encoding, each sensor makes decisions and
performs encoding upon each reception of a source packet. This mechanism significantly
reduces the node's storage requirements.

The remainder of this paper is organized as follows.  In
Sec.~\ref{sec:model}, we introduce the network and coding model. In
Sec.~\ref{sec:ltcds}, we present the LTCDS algorithm and provide its
performance analysis. In Sec.~\ref{sec:rcds}, we present the RCDS algorithm.
In Sec.~\ref{sec:sim}, we present simulation results for various performance measures of the proposed algorithms

\section{Network and Coding Models\label{sec:model}}

We model a wireless sensor network consisting of $n$ nodes as a random geometric graph~\cite{Gi61, Pe03}, as follows:
The nodes are distributed uniformly at random on the plane
and all have communication radii of 1. Thus, two nodes are neighbors and can communicate iff their distance is at most $1$.
Among the $n$ nodes, there are $k$ source nodes (uniformly and independently picked from the $n$)
that have independent information to be disseminated throughout the network for storage. A
similar model was considered in~\cite{LiLiLi07-1}. Our algorithms and results apply for many network topologies,
e.g., {regular grids of~\cite{DiPrRa05}.}

We assume that no node has knowledge about the locations of other nodes and no routing
table is maintained; thus the algorithm proposed in~\cite{DiPrRa05} cannot be
applied. Moreover, we assume that no node has
any global information, e.g., the total number of nodes $n$, the total number of
sources $k$, or the maximal number of neighbors in the network. Hence, the algorithms
proposed in~\cite{LiLiLi07-1} cannot be applied.
We assume that each node knows its neighbors. Let $\mathcal{N}(u)$
denote the set of neighbors of $u$. We will refer to the number of neighbors of $u$ as the \emph{node degree} of
$u$, and denote it by $\mu(u)=|\mathcal{N}(u)|$. The \emph{mean degree} of a graph $G$ is then given by
\begin{equation}\label{eq:mu}
\overline{\mu} = \frac{1}{|V|}\sum_{u\in G}\mu(u).
\end{equation}

For $k$ source blocks $\{x_1,\dots,x_k\}$ and a probability distribution $\Omega$ over the
set $\{1,\dots, k\}$, a Fountain code with parameters $(k,\Omega)$ is a potentially limitless stream of
output blocks $\{y_1,y_2,\dots\}$ \cite{Lu02,Sh06}. Each output block is generated by XORing $d$ randomly and
independently chosen source blocks, where $d$ is drawn from $\Omega(d)$.

LT (Luby Transform) codes \cite{Lu02,Sh06} are Fountain codes that employ either the
\emph{Ideal Soliton} distribution
\begin{equation}\label{eq:Ideal-Soliton-distribution}
\Omega_I(d)=\left\{ \begin{array}{ll} {1}/{k}, & d=1,\\
{1}/[{d(d-1)}], & d=2,3,\dots,k,\end{array}\right.
\end{equation}
or the \emph{Robust Soliton} distribution, which is defined as follows:
Let $R=c_0\ln(k/\delta)\sqrt{k}$, where $c_0$ is a suitable constant and $0<\delta<1$.
Define
\begin{equation}
\tau(d)=\left\{ \begin{array}{ll} {R}/{dk}, & d=1,\dots,{k}/{R}-1,\\
{R\ln(R/\delta)}/{k}, & d={k}/{R}, \\
0, & d={k}/{R}+1,\dots,k.\end{array}\right.
\end{equation}
The Robust Soliton distribution is given by
\begin{equation}\label{eq:Robust-Soliton-distribution}
\Omega_{R}(d)=\frac{\tau(d)+\Omega_{I}(d)}{\sum_{i=1}^k \big(\tau(i)+\Omega_{I}(i)\big)}, d=1,2,\dots,k.
\end{equation}
Raptor codes are concatenated codes whose inner codes are LT and outer codes are
traditional erasure correcting codes.
They have linear encoding and decoding complexity~\cite{Sh06}.

If each node in the network ends up storing an LT or Raptor code output
block corresponding to the $k$ source blocks, then the
the $k$ source blocks can be recovered in a computationally simple way from any
$k(1+\epsilon)$ of nodes for some small $\epsilon>0$, \cite{Lu02,Sh06}. For different
goals, different distributions $\Omega$ may be of interest. Our storage algorithm
can take any $\Omega$ as its input.


\section{LT Codes Based Algorithms\label{sec:ltcds}}

\subsection{Algorithm Design}
The goal of our storage algorithm is to have each of the $n$ nodes store an LT code output
block corresponding to the $k$ input (source) blocks without involvement of a central authority.
To achieve this goal, a node in a network would have to store, with probability $\Omega(d)$,
a binary sum (XOR) of $d$ randomly and independently chosen source packets. Our main idea to approach this goal
in a decentralized way is to (1) disseminate the $k$ source packets throughout the network by $k$ simple random walks
and (2) XOR a packet ``walking'' through a node with a probability $d/k$ where $d$ is chosen at the node
randomly according to $\Omega$.

To ensure that each of the $k$ random walks at least once visits each network node, we will let the random walks
last longer than the network (graph) cover time~\cite{AlFi02, Ro95}.
\begin{definition}\textit{(Cover Time)}
Given a graph $G$, let $T_{cover}(u)$ be the expected length of a simple random walk that starts at node
$u$ and visits every node in $G$ at least once. The \emph{cover time} of $G$ is defined by
$T_{cover}(G)=\max_{u\in G}T_{cover}(u)$.
\end{definition}
\begin{lemma}[{Avin and Ercal~\cite{AvEr05}}]\label{Lemma:Cover-Time}
Given a random geometric graph $G$ with $n$ nodes, if it is a connected graph with
high probability, then
\begin{equation}\label{eq:T-Cover-G}
T_{cover}(G)=\Theta(n\log n).
\end{equation}
\end{lemma}
In addition, the probability that a random walk on $G$ will require more time than $T_{cover}(G)$
to visit every node of $G$ is $\mathcal{O}({1}/{n\log n})$ \cite{AlFi02}.
Therefore, we can virtually ensure that a random walk visits each network node by requiring that it makes
$C_1n\log n$ steps for some $C_1>0$.
To implement this requirement for the $k$ random walks, we set a counter for each source packet and increment
it after each transmission.
Each time a node receives a packet whose counter is smaller than $C_1n\log n$,
it accepts the packet for storage with probability $d/k$ (where $d$ is chosen at the node according to
$\Omega$), and then, regardless of
the acceptance decision, it forwards the packet to one of its randomly chosen neighbors.
Packets older than $C_1n\log n$ are discarded.

Note that the above procedure requires the knowledge of $n$ and $k$ at each node. To devise a fully
decentralized storage algorithm, we note that
each node can observe (1) how often it receives a packets and (2) how often it receives
a packets from each source. Naturally, one expects that these numbers depend on the network
connectivity ($\mu(u)$ for all $u$), the size of the graph $n$, and the number
of different random walks $k$. We next describe this dependence and show how it can be
used to obtain local estimates of global parameters.

The following definitions and claims either come from~\cite{AlFi02,Ro95,MoRa95}, or can be easily derived
based on the results therein.
\begin{definition}\textit{(Inter-Visit Time)}
For a random walk on a graph, the \emph{inter-visit time} of node $u$, $T_{visit}(u)$, is the
amount of time between any two consecutive visits of the  walk to $u$.
\end{definition}
\begin{lemma}\label{Lemma:Inter-Visit-Time}
For a node $u$ with node degree $\mu(u)$ in a random geometric graph, the mean inter-visit time is
\begin{equation}\label{eq:E-T-visit-u}
E[T_{visit}(u)]={\overline{\mu} n}/{\mu(u)},
\end{equation}
where $\overline{\mu}$ is the mean degree of the graph given by~\eqref{eq:mu}.
\end{lemma}

Lemma~\ref{Lemma:Inter-Visit-Time} implies $n= {\mu(u)E[T_{visit}(u)]}/{\overline{\mu}}$.
While node $u$ can easily measure $E[T_{visit}(u)]$, the mean degree $\overline{\mu}$ is a piece of global information
and may be hard to obtain.
Thus we make a further approximation and let the estimate of $n$ by node $u$ be
\begin{equation}
\hat{n}(u) = E[T_{visit}(u)].
\label{eq:n}
\end{equation}

Note that to estimate $n$, it is enough to consider only one of the $k$ random walks.
Now to estimate $k$, we also need to consider the $k$ walks jointly without distinguishing between packets originating
from different sources.
\begin{definition}\textit{(Inter-Packet Time)}
For multiple random walks on a graph, the \emph{inter-packet time} of node $u$, $T_{packet}(u)$, is the
amount of time between any two consecutive visits by any of the walks to $u$.
\end{definition}
\begin{lemma}\label{Lemma:Inter-Packet-Time}
For a node $u$ with node degree $\mu(u)$ in a random geometric graph with $k$ simple random walks,
the mean inter-packet time is
\begin{equation}\label{eq:E-T-packet-u}
E[T_{packet}(u)]=\frac{E[T_{visit}(u)]}{k}=\frac{\overline{\mu} n}{k\mu(u)},
\end{equation}
where $\overline{\mu}$ is the mean degree of the graph given by~\eqref{eq:mu}.
\end{lemma}
\emph{Proof:} For a given node $u$, each of the $k$ random walks
has expected inter-visit time $\frac{\overline{\mu} n}{\mu(u)}$. We now view this process from
another perspective: we assume there are $k$ nodes $\{v_1,\dots,v_k\}$ uniformly
distributed in the network and an agent from node $u$ following a simple random walk. Then
the expected inter-visit time for this agent to visit any particular $v_i$ is the same as
$\frac{\overline{\mu} n}{\mu(u)}$. However, the expected inter-visit time for any two nodes $v_i$
and $v_j$ is
$
\frac{1}{k}\frac{\overline{\mu} n}{\mu(u)}$,
which gives (\ref{eq:E-T-packet-u}).\qed
Based on Lemmas~\ref{Lemma:Inter-Visit-Time} and \ref{Lemma:Inter-Packet-Time},
that is equations (\ref{eq:E-T-visit-u}) and (\ref{eq:E-T-packet-u}), we see
that each node $u$, can estimate $k$ as
\begin{equation}
\hat{k}(u)={E[T_{visit}(u)]}/{E[T_{packet}(u)]}.
\end{equation}
We are now ready to state the entire storage algorithm:
\newpage
\begin{definition}\textit{(LTCDS Algorithm)}\\ with system parameters $C_1,C_2>0$ and $\Omega$\\[1mm]
{\bf Initialization Phase}\\
Each source node $s, s=1,\dots,k$
  \begin{enumerate}
\item  attaches a header to its data $x_{s}$, containing
its ID and a life-counter $c(x_{s})$ set to zero, and then
\item sends its packet to a randomly selected neighbor.
\end{enumerate}
Each node $u$ sets its storage $y_u = 0$.\\
{\bf Inference Phase} (at all nodes $u$)
  \begin{enumerate}
\item Suppose $x_{s(u)_1}$ is the first source packet that visits $u$, and
denote by $t_{s(u)_1}^{(j)}$ the time when $x_{s(u)_1}$ makes its $j$-th visit to $u$.
Concurrently, $u$ maintains a record of visiting times for all packets
$x_{s(u)_i}$ ``walking'' through it. Let $t_{s(u)_i}^{(j)}$ be the time when source packet $x_{s(u)_i}$
makes its $j$-th visit to $u$. After $x_{s(u)_1}$ visits $u$ $C_2$ times, where
$C_2>0$ is system parameter, $u$ stops this monitoring and
recoding procedure. Denote by $k(u)$ the number of source packets that have visited at least once
until that time.
\item Let $J(s(u)_i)$ be the number of visits of source packet $x_{s(u)_i}$ to $u$ and let
    \begin{eqnarray}\label{eq:T-s-u-i}
    T_{s(u)_i}&=& \frac{1}{ {J(s(u)_i)-1}} \sum_{j=1}^{ {J(s(u)_i)-1}}\Bigl(t_{s(u)_i}^{(j+1)}-t_{s(u)_i}^{(j)}\Bigr) \nonumber\\
    &=&\frac{1}{ {J(s(u)_i)-1}} \Bigl( t_{s(u)_i}^{(  J(s(u)_i)  )}-t_{s(u)_i}^{(1)}\Bigr).
    \end{eqnarray}
    Then, the average inter-visit time for node $u$ is
    \begin{equation}\label{eq:bar-T-visit-u}
    \bar{T}_{visit}(u)= \frac{1}{k(u)} \sum_{i=1}^{k(u)}T_{s(u)_i}.
    \end{equation}
     \[ \text{Let} ~ J_{min}=\min_{s(u)_i}\{t_{s(u)_i}^{(1)}\} ~ \text{and} ~
J_{max}=\max_{s(u)_i}\{t_{s(u)_i}^{(J(s(u)_i))}\}.
\] Then the inter-packet time is
    \begin{equation}\label{eq:bar-T-packet-u}
    \bar{T}_{packet}(u)= \frac{ {J_{max}-J_{min}}}{\sum_{s(u)_i}J(s(u)_i)},
    \end{equation}
    and $u$ can estimate $n$ and $k$ as
    \begin{equation}\label{eq:n-tilde}
    \hat{n}(u)=\bar{T}_{visit}(u)
    ~~ \text{and} ~~
    \hat{k}(u)=\frac{\bar{T}_{visit}(u)}{\bar{T}_{packet}(u)}.
    \end{equation}
\item In this phase, the counter $c(x_{s_i})$ of each source packet $c(x_{s_i})$ is incremented by
one after each transmission.
\end{enumerate}
{\bf Encoding and Storage Phase} (at all nodes $u$)
\begin{enumerate}
\item Node $u$ draws $d_c(u)$  from $\{1,\dots,\hat{k}(u)\}$ according to $\Omega$.
\item Upon reciving packet $x$, if $c(x)< C_1\hat{n}\log \hat{n}$, node $u$
\begin{itemize}
  \item puts $x$ into its forward queue and increments $c(x)$.
  \item with probability ${d_c(u)}/{\hat{k}}$, accepts $x$ for storage and updates its storage variable $y_u^-$
  to $y_u^+$ as
    \begin{equation}
    y_u^+ = y_u^- \oplus x_{s},
    \end{equation}
\end{itemize}
If $c(x)< C_1\hat{n}\log \hat{n}$, $x$ is removed from circulation.
\item When a node receives a packet before the current round, it
forwards its head-of-line (HOL) packet to a randomly chosen neighbor.
\item
Encoding phase ends and storage phase begins when each node has seen its $\hat{k}(u)$ source packets.
\end{enumerate}
\end{definition}

\subsection{Performance Analysis}
Parameters $(k,\Omega)$ determine the error rate performance and encoding/decoding complexity
of the corresponding Fountain code. With input $(k,\Omega)$, the LTCDS algorithm produces a
distributed Fountain code with parameters $(k,\Omega^{\prime})$, where
$\Omega^{\prime}\ne \Omega$. We next compute $\Omega^{\prime}$ when the
input distribution $\Omega$ is the Robust Soliton (\ref{eq:Robust-Soliton-distribution}),
and discuss the performance and complexity of the corresponding Fountain code.

Recall that node $u$ draws ${d_c(u)}$ according to $\Omega$, and accepts a passing
source packet with probability ${d_c(u)}/{k}$. Therefore,
the number of packets that $u$ accepts, given $d_c(u)$, is Binomially distributed with
parameter ${d_c(u)}/{k}$, and the number of packets that $u$ accepts takes value $i$
with probability $\Omega^{\prime}(i)$:
\begin{align*}
\Omega^{\prime}(i)&=\sum_{d_c(u)=1}^{k}\Pr(\tilde{d}_c(u)=i|d_c(u))\Omega(d_c(u))\nonumber\\
& = \sum_{d_c(u)=1}^{k}
\binom{k}{i}\left(\frac{d_c(u)}{k}\right)^i\left(1-\frac{d_c(u)}{k}\right)^{k-i}\Omega(d_c(u)).
\end{align*}

A simple way to achieve $\Omega^{\prime} = \Omega$
would be to let each $u$ store each distinct passing source packet until it collects all $k$,
and then randomly choose exactly $d_c(u)$ packets, where ${d_c(u)}$ is drawn according to $\Omega$,
This approach would require large buffers, which is usually not practical, especially when $k$ is
large. Therefore, we assume that nodes have limited memory and let them make their decision
upon each reception. Our approach, as the following theorem shows, results in a Fountain code with
comparable efficiency and the same complexity as the one determined by the Robust or Ideal Soliton
distributions.

\begin{theorem}\label{Theorem:Decoding-LTCDS-I}
Suppose the LTCDS algorithm uses the Robust Soliton distribution (\ref{eq:Robust-Soliton-distribution}) for
$\Omega$. Then, the $k$
source packets can be recovered from any $K^\prime=\beta K$ nodes with probability $1-\delta$
for sufficiently large $k$, where $\beta\geq (1-e^{-2})^{-1}$ and
$K=k+\mathcal{O}\big(\sqrt{k}\log^2(k/\delta)\big)$
($K$ would be sufficient for recovery when  $\Omega^{\prime} = \Omega$). The decoding complexity is $\mathcal{O}(k\log(k/\delta))$.
\end{theorem}

\emph{Proof:}  {The probability that a node stores no information is}
\begin{align}\label{eq:P-0-1}
\Omega^{\prime}(0)&=
\sum_{d=1}^k \left(1-\frac{d}{k}\right)^k\Omega(d)< \sum_{d=1}^k e^{-d}\Omega(d)\nonumber\\
&< \sum_{d=1}^k\tau(d)e^{-d} +\sum_{d=1}^k\Omega_I(d)e^{-d}\nonumber\\
&= \sum_{d=1}^{\frac{k}{R}-1}\frac{R}{kd}e^{-d}+\frac{R\ln
(\frac{R}{\delta})}{k}e^{-k/R}+\sum_{d=1}^k\Omega_I(d)e^{-d}\nonumber\\
&< \frac{R}{k}\sum_{d=1}^{\frac{k}{R}-1}\frac{e^{-1}}{d}+\frac{R\ln
(\frac{R}{\delta})e^{-\frac{k}{R}}}{k}+\frac{e^{-1}}{k}+\sum_{d=2}^k\frac{e^{-2}}{d(d-1)}\nonumber\\
&<\mathcal{O}\left(\frac{(\ln k)^2}{\sqrt{k}}\right)+e^{-2}.
\end{align}
Therefore, for sufficiently large $k$, $\Omega^{\prime}(0)<e^{-2}$. Consequently, if we randomly take
$K^{\prime}=\beta K$ nodes from the network, where $\beta\geq (1-e^{-2})^{-1}$, we have
\begin{align*} \Pr\bigl\{N_0<(1-\alpha)&K^{\prime}\bigl(1-\Omega^{\prime}(0)\bigr)\bigr\}\leq   \\
&\frac{K^{\prime}\Omega^{\prime}(0)\bigl(1-\Omega^{\prime}(0)\bigr)}{\alpha^2K^{\prime 2}\bigl(1-\Omega^{\prime}(0)\bigr)^2}
=\Theta\left(\frac{1}{k}\right),
\end{align*} for any $\alpha>0$, where $N_0$ denotes the number of nodes that store encoded packets. Therefore,
we have $K^{\prime}(1-e^{-2})\geq K$ nodes that store encoded packets with a high probability for
sufficiently large $n$ and $k$.

We next show that the original $k$ source packets can be recovered based on
$K=k+\mathcal{O}\big(\sqrt{k}\log^2(k/\delta)\big)$ stored packets with
probability $1-\delta$, by an argument very similar to the one in~\cite{Lu02}. When a source packet is decoded (e.g., from stored packets with degree one), we say that all the
other encoded packets that contain this source packet are covered. In the decoding process, call
the set of covered encoded packets that have not been fully decoded (all the contained source
packets are decoded) as the ripple. The main idea of the proof is to show the ripple size
variation is very similar to a random walk, and the probability that the ripple size deviates from
its mean in $k$ steps by $\Theta(\sqrt{k})$ is small~\cite{Lu02}.

It can be shown that the expected number of stored packets of degree one is $\theta' R$ for some
constant $\theta'>0$. Employing a Chernoff bound argument, we can show that with probability at
least $1-\delta/3$, the initial ripple size due to degree one packets is at least $\theta R/2$ for
a suitable constant $\theta>0$. Then by the same argument used in the proof for Theorem 17
in~\cite{Lu02}, it can be shown that without contribution of $\tau(k/R)$ in $\Omega$, the
ripple does not disappear for $L=k-1,\dots,R$ and the decoding process is successful until $R$
stored packets remain undecoded with probability at least $1-\delta/3$.

Further, like Proposition 15 in~\cite{Lu02}, we can show that using only the
contribution of $\tau(k/R)$ in $\Omega$, the last $R$ blocks can be decoded with probability $1-\delta/3$
when between $2R$ and $R$ stored packets remain undecoded . This implies that the
decoding process completes successfully with probability $1-\delta$.

Finally, the decoding complexity is the average degree $D$ of a stored packet:
\begin{eqnarray}
D\!\!\!\!\!&=&\!\!\!\!\! \sum_{i=1}^k i\left[\sum_{d=1}^{k}
\binom{k}{i}\left(\frac{d}{k}\right)^i\left(1-\frac{d}{k}\right)^{k-i}\Omega(d)
\right]\nonumber\\
\!\!\!\!\!&=&\!\!\!\!\!\sum_{d=1}^k k\left[\sum_{i=1}^{k}
\binom{k-1}{i-1}\left(\frac{d}{k}\right)^i\left(1-\frac{d}{k}\right)^{k-i}\right]\Omega(d)\nonumber\\
\!\!\!\!\!&=&\!\!\!\!\!\sum_{d=1}^k d\left[\sum_{i=0}^{k-1}
\binom{k-1}{i}\left(\frac{d}{k}\right)^{i}\left(1-\frac{d}{k}\right)^{k-1-i}\right]\Omega(d)\nonumber\\
\!\!\!\!\!&=&\!\!\!\!\!\sum_{d=1}^k d\Omega(d)=\mathcal{O}(\log{k/\delta})
\end{eqnarray}
where the last equality is due to Theorem 13 in~\cite{Lu02}. \qed

From the calculation of $\Omega^{\prime}(0)$, with the Robust or Ideal Soliton distribution, we also have
\begin{equation}\label{eq:P-0-2}
\Omega^{\prime}(0)>\frac{1}{2e^{2}}.
\end{equation}

 {\emph{Remark:} One interesting implication of~\eqref{eq:P-0-1}
and~\eqref{eq:P-0-2} is that in order to achieve the same performance as that of original LT
codes, more than $(1-e^{-2}/2)^{-1}K\approx 1.07K$ nodes, but less than $(1-e^{-2})^{-1}K\approx
1.15K$ nodes are required to recover the original $k$ source packets.}

Another main performance metric is the transmission cost of the algorithm, which is characterized
by the total number of transmissions (the total number of steps of $k$ random walks).
\begin{theorem}\label{Theorem:Transmission-LTCDS-II}
The total number of transmissions of the LTCDS algorithm is $\Theta(kn\log n)$.
\end{theorem}
\emph{Proof:} In the interference phase of the LTCDS algorithm, the total number of transmissions
is upper bounded $C'n$ for some constant $C'>0$. That is because each node needs to receive the
first visit source packet for $C_2$ times, and by Lemma~\ref{Lemma:Inter-Visit-Time}, the mean
inter-visit time is $\Theta(n)$.
In the encoding phase, in order to guarantee that each source packet visits all the nodes,
the number of steps of each of the $k$ random walks is required to be $\Theta(n\log n)$. Since there are $k$ source packets,
the total number of transmissions algorithm is $\Theta(kn\log n)$.\qed

Note that the algorithm proposed in~\cite{LiLiLi07-1} has similar order of total number transmissions. If geometric information is available, as in~\cite{DiPrRa06-2}, the complexity can
be reduced, e.g., $\Theta(k\sqrt{n}\log n)$ for the algorithm proposed in~\cite{DiPrRa06-2}.

\section{Raptor Codes Based Algorithms\label{sec:rcds}}
Recall that Raptor codes are concatenated codes whose inner codes are LT and outer codes
(pre-codes) are traditional erasure correcting codes.
For the pre-codes will use is randomized LDPC codes with $k$ inputs and $m$ outputs ($m\geq k$).
Assume $n$ and $k$ are known or have been estimated at every node. To perform the LDPC coding for $k$
sources in a distributed manner, we again use simple random walks. Each
source node first generates $b$ copies of its own source packet, where $b$ follows some distribution
$P_{\text{LDPC}}$ defining the LDPC precode. (See \cite{Sh06} for the design of randomized LDPC
codes for Raptor codes.) These $b$ copies are then sent into the network by random walks.
Each of the remaining $n-k$ nodes in the network chooses to serve
as a parity node with probability $(m-k)/(n-k)$. We refer to the parity nodes together with the
original (systematic) source nodes as the pre-coding output nodes. All pre-coding output nodes accept a
source packet copy with the same probability; consequently, the $b$ copies of a given source packet
get distributed uniformly among all pre-coding output nodes. In this way, we have $m$
pre-coding output nodes, each of which contains an XOR of a random number of
source packets. The detailed description of the pre-coding algorithm is given below.
After obtaining the $m$ pre-coding outputs, to obtain Raptor codes based distributed storage,
we  apply the LTCDS algorithm with these $m$
nodes as new sources and an appropriate $\Omega$ as discussed in \cite{Sh06}.

\begin{definition}
\textit{(Pre-coding Algorithm)}
\begin{enumerate}
\item Each source node $s, s=1,\dots,k$ draws a random number $b(s)$ according to the
distribution of  {predefined LDPC codes}, generates $b(s)$ copies of
its source packet $x_{s}$ with its ID and a counter
$c(x_{s})$ with an initial value of zero in the packet header, and sends each of them to
one of its randomly chosen neighbors.
\item Each of the remaining $n-k$ nodes chooses to serve as a parity
node with probability $({m-k})/({n-k})$.  {These} parity nodes and the
original source nodes  {are} pre-coding output nodes. Each pre-coding
output node $w$ generate a random number $a(w)$ according to the following distribution: \[
\Pr(a(w)=d) = \binom{k}{d}\left(\frac{E[b]}{m}\right)^d \left( 1-\frac{E[b]}{m}\right)^{k-d},
\] where $E[b]=\sum_b bP_{\text{LDPC}}(b)$.
\item Each node that has packets in its forward queue before the current round sends its
HOL packet to one of its randomly chosen neighbors.
\item When a node $u$ receives a packet $x$ with $c(x)<C_3n\log(n)$, $u$
puts the packet into its forward queue and increments the counter.
\item Each pre-coding output node $w$ accepts the first $a(w)$ copies of different $a(w)$ source
packet with counters  {$c(x)\geq C_3n\log(n)$}, and updates $w$'s pre-coding
result each time as
    \begin{equation}
     {y_w^+ = y_w^- \oplus x}.
    \end{equation}
If a copy of $x$ is accepted, it will not be forwarded any more, and $w$ will
not accept any other copy of $x_{s_j}$. When the node $w$ completes $a(w)$ updates, $y_w$
becomes its pre-coding packet.
\end{enumerate}
\end{definition}

\section{Performance Evaluation\label{sec:sim}}

We evaluate the performance of LTCDS and RCDS algorithms by simulation. Our main performance
metric is the successful decoding probability vs.\ the query ratio.
\begin{definition} The
\emph{query ratio} $\eta$ is the ratio between the number of queried nodes $h$ and the
number of sources $k$:
\begin{equation}\label{eq:eta}
\eta={h}/{k}.
\end{equation}
\end{definition}

\begin{definition}\textit{(successful decoding)} We say that decoding is
\emph{successful} if it results in recovery of \textit{all} $k$ source packets.
\end{definition}

For a query ratio $\eta$, we evaluate $P_s$ by simulation as follows:
Let $h=\eta k$ denote the number of queried nodes. We select (uniformly at random) $10\%$ of
the $\binom{n}{h}$ possible subsets of size $h$ of the $n$ network nodes, and try to decode
the $k$ source packets from each subset. Then the fraction of times the decoding is successful
measures our $P_s$.

\begin{figure}[hbt]
\centerline{
\includegraphics[scale=0.4]{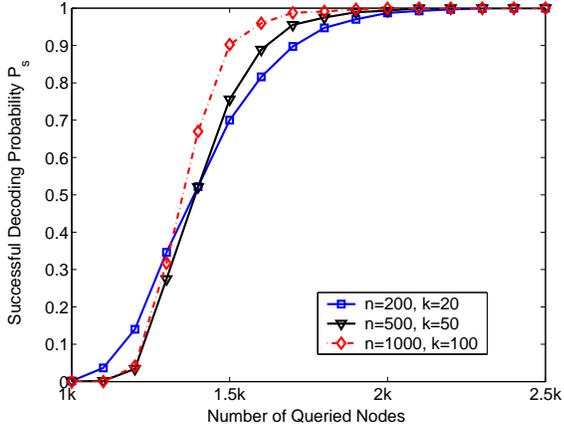}}
\caption{Performance of LTCDS with known $n$ and $k$ for (a) $n$=200, $k$=20; (b) $n$=500, $k$=50; and (c)
$n$=1000, $k$=100.}\label{fig:Encoding-I}
\end{figure}

Fig.~\ref{fig:Encoding-I} shows the decoding performance of LTCDS algorithm with known
$n$ and $k$.
For $\Omega$, we chose the Ideal Soliton distribution (\ref{eq:Ideal-Soliton-distribution}). The network is deployed in
$\mathcal{A}=[0,5]^2$ with density $\lambda=\frac{40}{9}$, and the system parameter $C_1=3$.
From the simulation results, we can see that when the query ratio is
above 2, the successful decoding probability $P_s$ is about $99\%$. When $n$ increases
but $k/n$ and $\eta$ remain constant, $P_s$
increases when $\eta\geq 1.5$ and decreases when $\eta<1.5$. {This} is
because when there are more nodes,  {it is more likely that} each node has
the Ideal Soliton distribution.

\begin{figure}[hbt]
\centerline{
\includegraphics[scale=0.4]{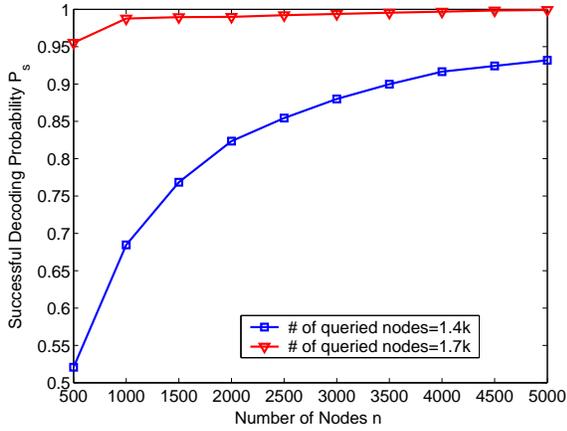}}
\caption{Performance of LTCDS with different known $n$ and $k$ and fixed number
of queried nodes for two cases: (a) $\eta=1.4$; (b) $\eta=1.7$.}\label{fig:Encoding-II}
\end{figure}

In Fig.~\ref{fig:Encoding-II},
we fix $\eta$ to 1.4 and 1.7 and $k/n=0.1$. From the results, it can be seen that as $n$
 {increases}, $P_s$ increases until it reaches a plateau, which is the
successful decoding probability of LT codes.

\begin{figure}[hbt]
\centerline{
\includegraphics[scale=0.4]{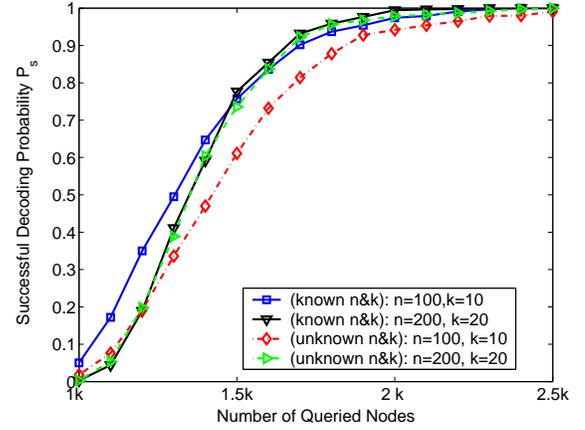}}
\caption{Performance of LTCDS algorithm with small number of nodes and sources
for (a) known $n$=100 and $k=10$; (b) known $n$=200 and $k=20$; (c)
unknown $n$=100 and $k=10$; (d) unknown $n$=200 and $k=20$.}\label{fig:LTCDS-I}
\end{figure}

\begin{figure}[hbt]
\centerline{
\includegraphics[scale=0.4]{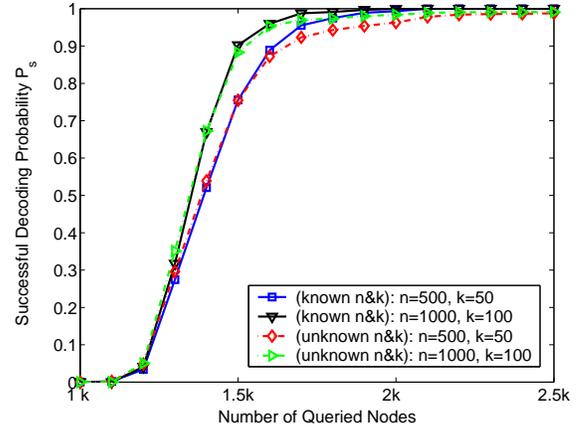}}
\caption{Performance of LTCDS algorithm with large number of nodes and sources
for (a) known $n$=500 and $k=50$; (b) known $n$=1000 and $k=100$; (c)
unknown $n$=500 and $k=50$; (d) unknown $n$=1000 and $k=100$.}\label{fig:LTCDS-II}
\end{figure}

We compare the decoding performance of LTCDS with known and unknown values of $n$ and $k$
in Fig.~\ref{fig:LTCDS-I} and Fig.~\ref{fig:LTCDS-II}.
The network is
deployed in $\mathcal{A}=[0,5]^2$, and the system parameter is set as $C_1=10$. To
guarantee each node to obtain accurate estimates of $n$ and $k$, we set $C_2$ large
enough as $C_2=50$. The decoding performance of the LTCDS algorithm with unknown $n$ and
$k$ is a little bit worse than that of the LTCDS algorithm with known $n$ and $k$ when
$\eta$ is small, and almost the same when $\eta$ is
large. Such difference between the two algorithms becomes marginal when the number of
nodes and sources increase as shown in Fig.~\ref{fig:LTCDS-II}.

An interesting observation in Fig.~\ref{fig:Encoding-I}, Fig.~\ref{fig:LTCDS-I} and
Fig.~\ref{fig:LTCDS-II} is that the probability of successful decoding is almost zero
until we query about $1.1k$ nodes. This is due to the nodes that store no information in
the network. As we pointed out in the Remark after the proof of
Theorem~\ref{Theorem:Decoding-LTCDS-I}, for Robust Soliton distribution, more than
$1.07k$ but less than $1.15k$ nodes are needed to query to achieve the same performance
of LT codes.  {Similar} results also hold for Ideal Soliton distribution.

\begin{figure}[hbt]
\centerline{
\includegraphics[scale=0.38]{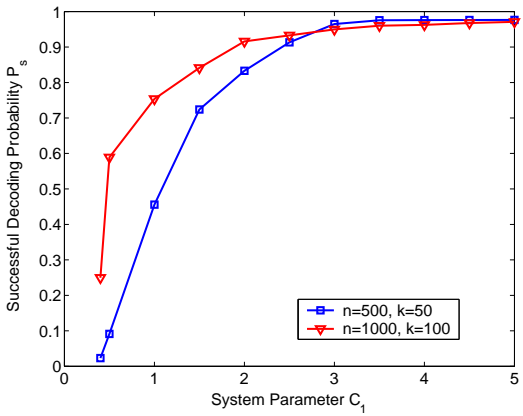}}
\caption{Performance of LTCDS algorithm with different system parameter $C_1$
for two cases: (a) $n=500$ and $k=50$, (b) $n=1000$ and
$k=100$.}\label{fig:Parameters-I}
\end{figure}

To investigate how the system parameter $C_1$ affects the decoding performance of the
LTCDS algorithm with known $n$ and $k$, we fix $\eta$ and vary $C_1$. The simulation
results are shown in Fig.~\ref{fig:Parameters-I}.
When $C_1\geq 3$, $P_s$ keeps almost
like a constant, which indicates that after $3n\log n$ steps, almost all source packets
visit each node at least once.

\begin{figure}[hbt]
\centerline{
\includegraphics[scale=0.38]{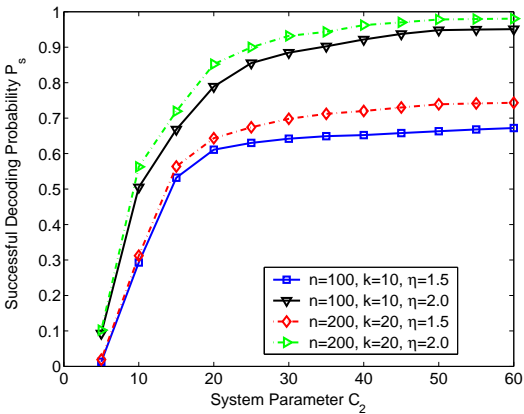}}
\caption{Performance of LTCDS algorithm with different system parameter $C_2$
for (a) $n=100$, $k=10$, $\eta=1.5$; (b) $n=100$, $k=10$, $\eta=2.0$;
(c) $n=200$, $k=20$, $\eta=1.5$; (d) $n=200$, $k=20$,
$\eta=2.0$.}\label{fig:Parameters-II}
\end{figure}

Furthermore, to investigate how the system parameter $C_2$ affects the decoding
performance of the LTCDS algorithm, we fix $\eta$ and $C_1$, and vary $C_2$. From
Fig.~\ref{fig:Parameters-II}, we can see that when $C_2$ is small, the performance of the
LTCDS algorithm is very poor. This is due to the inaccurate estimates of $k$ and $n$ by
each node. When $C_2$ is large, for example, when $C_2\geq 30$, the performance is almost
the same.

\begin{figure}[hbt]
\centerline{ \subfigure[]{
\includegraphics[scale=0.23]{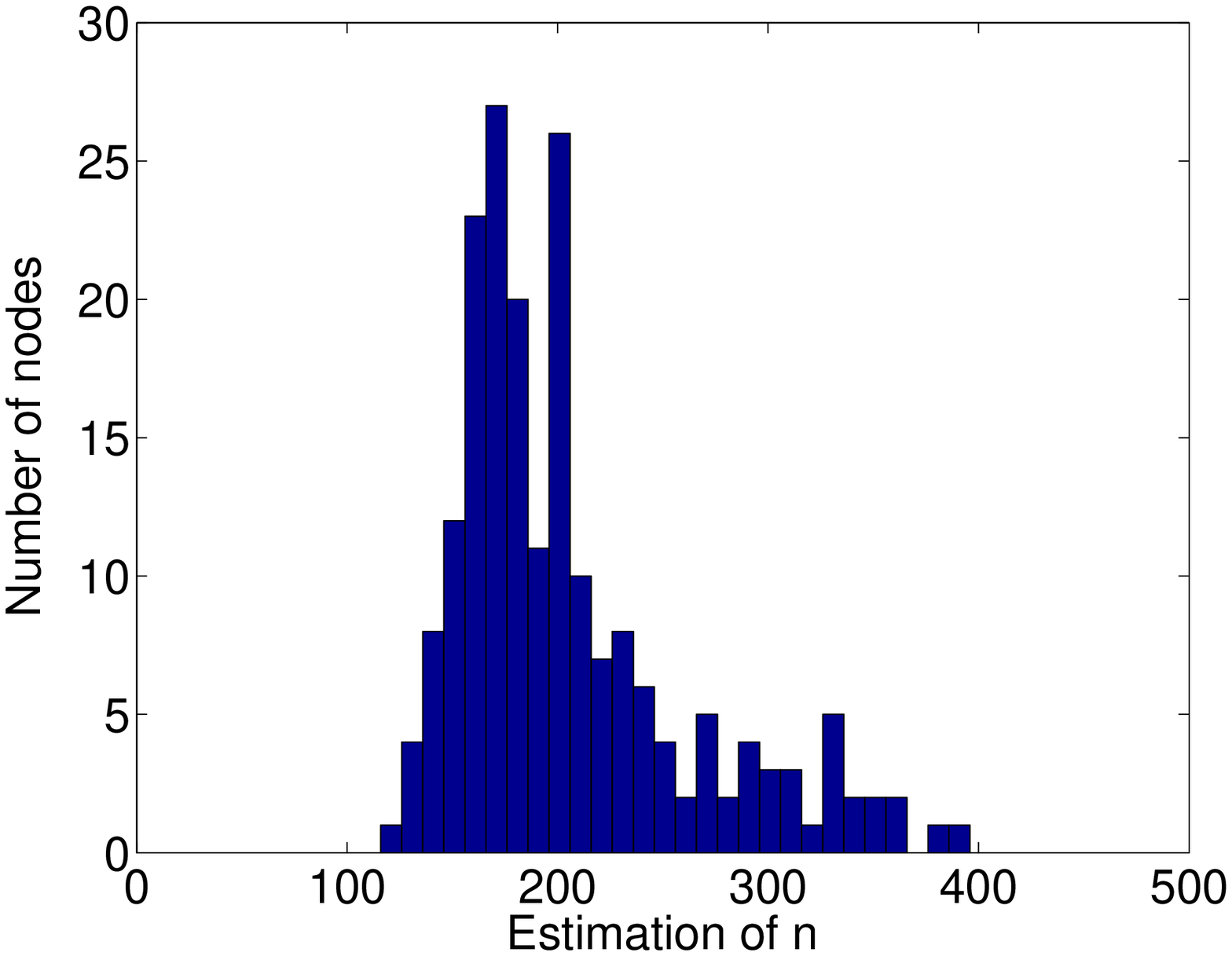}}\hfil
\subfigure[]{
\includegraphics[scale=0.23]{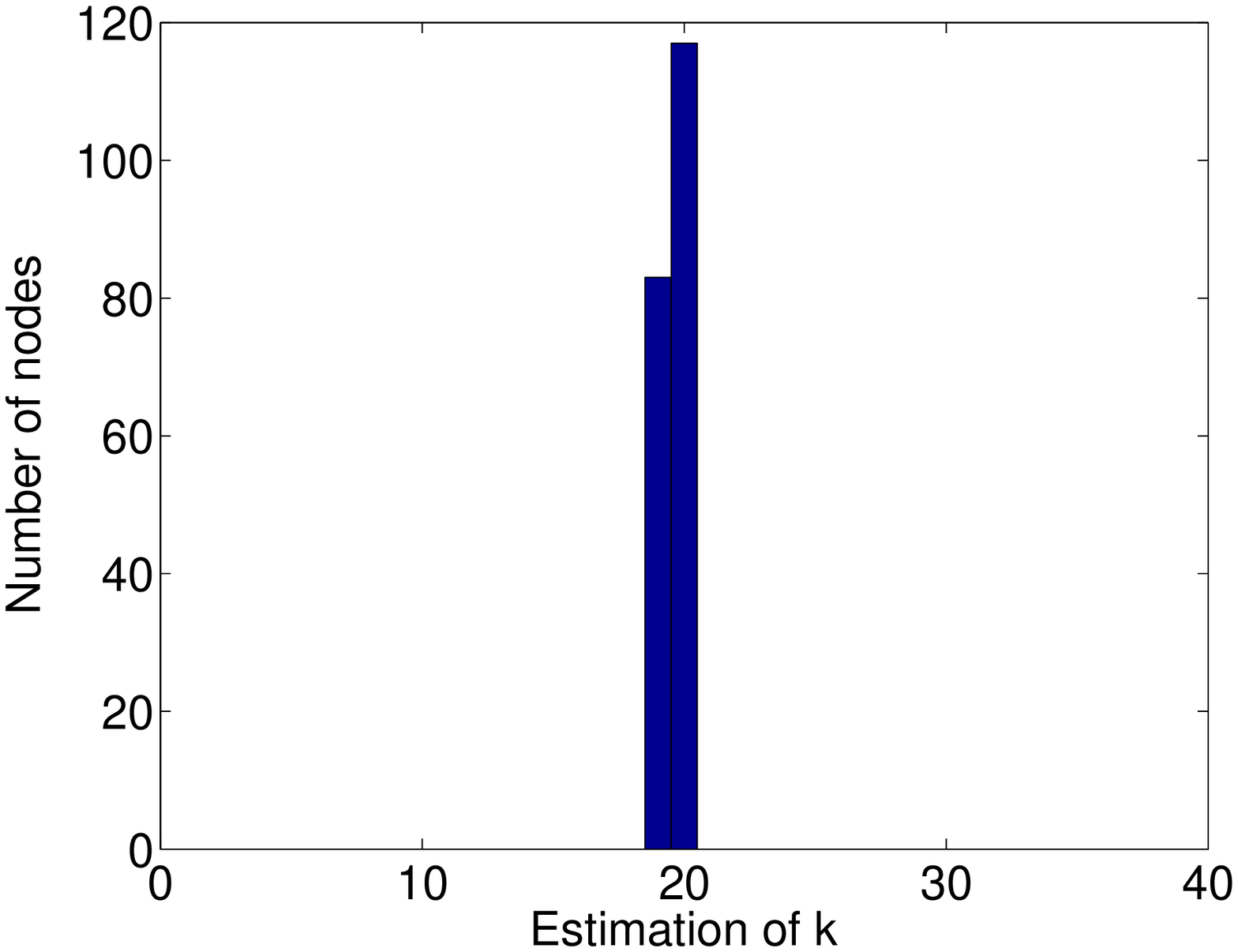}}}
\caption{Histograms for estimates of $n$ (a) and $k$ (b) in LTCDS algorithm with $n=200$ and $k=20$.}
\label{fig:LTCDS-II-Est-2}
\end{figure}

\begin{figure}[h]
\centerline{ \subfigure[]{
\includegraphics[scale=0.23]{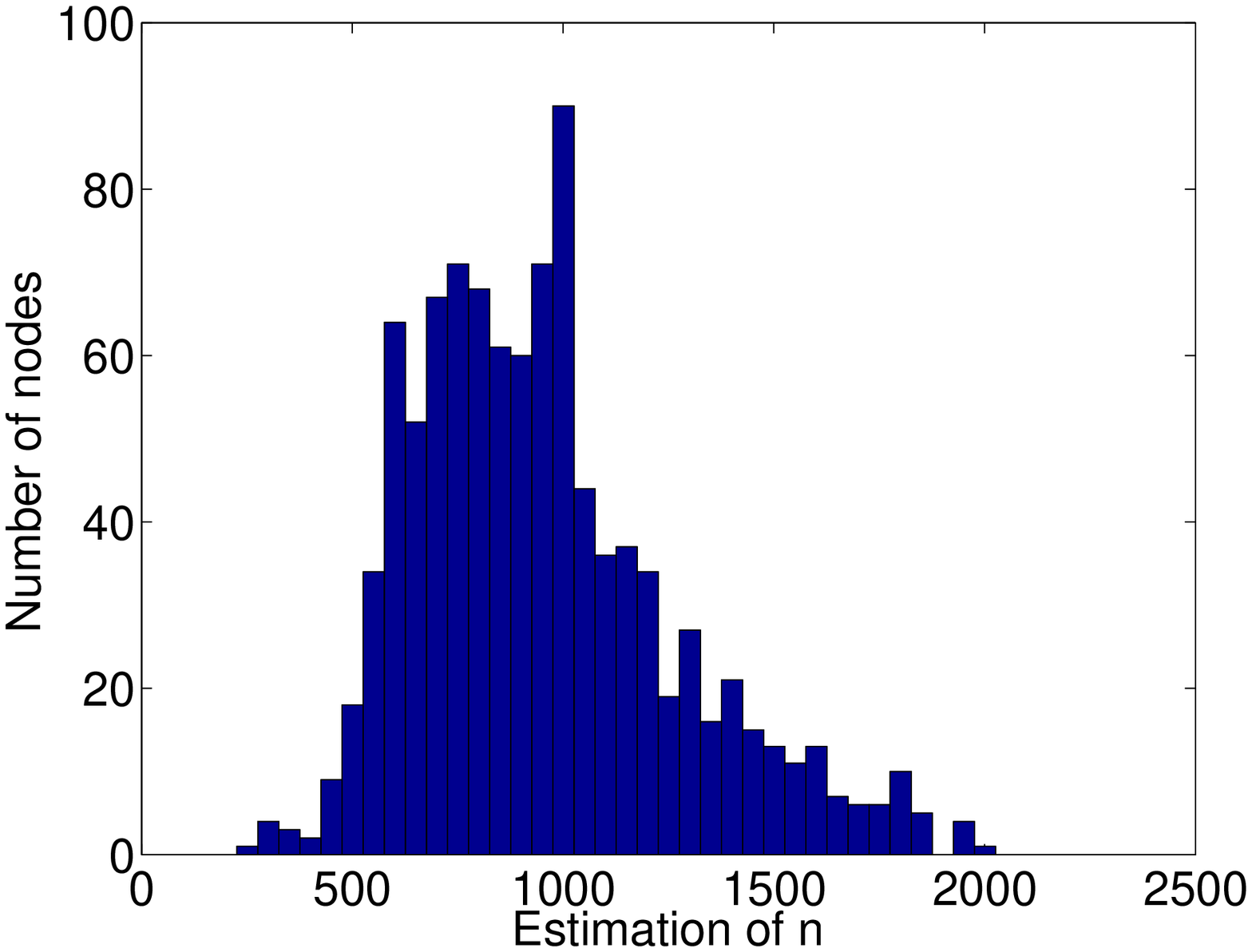}}\hfil
\subfigure[]{
\includegraphics[scale=0.23]{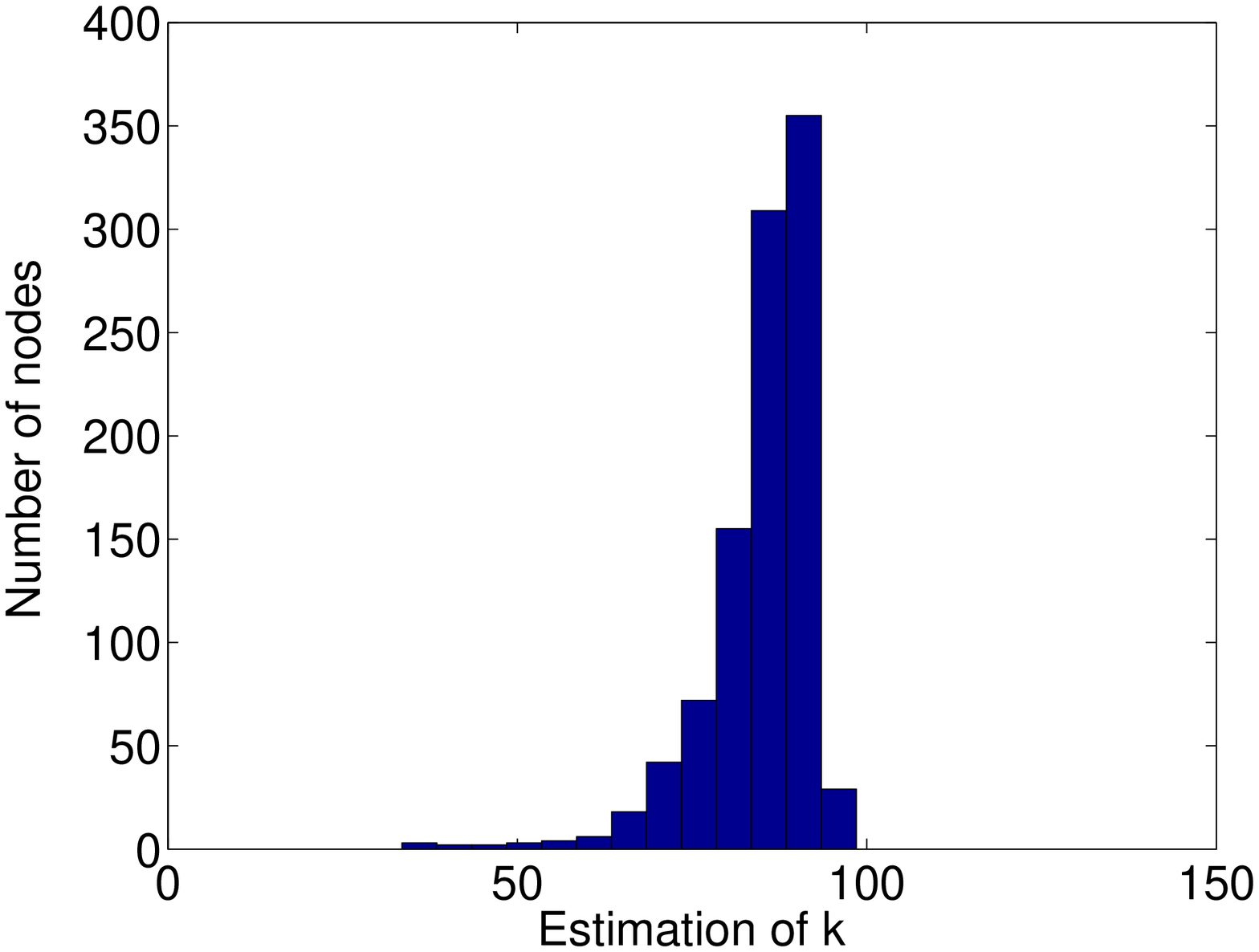}}}
\caption{Histograms for estimates of $n$ (a) and $k$ (b) in LTCDS algorithm with $n=1000$ and $k=100$.}
\label{fig:LTCDS-II-Est-3}
\end{figure}

Fig.~\ref{fig:LTCDS-II-Est-2} and Fig.~\ref{fig:LTCDS-II-Est-3} {show}
the histograms of the estimation results for $n$ and $k$, based on equations
(\ref{eq:n-tilde}).
As expected, the estimates of $k$ are more accurate and concentrated than
the estimates of $n$.

\bibliographystyle{ieeetr}
\end{document}